\documentclass[11pt, a4paper]{article}
\usepackage{moriond,epsfig}
\usepackage{wrapfig}
\usepackage{graphicx}
\usepackage{psfig}

\def\be{\begin{equation}}
\def\ee{\end{equation}}
\def\bea{\begin{eqnarray}}
\def\eea{\end{eqnarray}}

\def\msol{{M_\odot}}
\def\lsb{L_{\rm SB}}

\begin{document}
\vspace*{4cm}
\title{Keynote Lecture: Galactic and Extragalactic Bubbles}

\author{ D. BREITSCHWERDT (1), M.A. DE AVILLEZ (1,2), M.J. FREYBERG (3)}
\address{(1) Institut f\"ur Astronomie, Universit\"at Wien,
        T\"urkenschanzstra{\ss}e 17, A-1180 Wien, Austria}
\address{(2) Department of Mathematics, University of \'Evora,
              R. Rom\~ao Ramalho 59, 7000 \'Evora, Portugal}
\address{(3) MPI f\"ur extraterrestrische Physik,
Giessenbachstra{\ss}e 1, D-85748 Garching, Germany}

\maketitle\abstracts{ The observational and theoretical state of
Galactic and extragalactic bubbles are reviewed. Observations of
superbubbles are discussed, with some emphasis on nearby bubbles
such as the Local Bubble (LB) and the Loop~I superbubble (LI).
Analytical bubble theory is revisited, and similarity solutions,
including the time-dependent energy input by supernova explosions
according to a Galactic initial mass function (IMF), are studied.
Since the agreement with observations is not convincing in case of
the LB, we present high resolution 3D AMR simulations of the LB and
LI in an inhomogeneous background medium. It is demonstrated that
both the morphology and recently published FUSE data on O{\sc vi}
absorption line column densities can be well understood, if the LB
is the result of about 20 supernova explosions from a moving group,
and the LB age is about 14.7 Myrs. } %
\noindent
{\small{\it Keywords}: ISM: general - ISM: evolution -
ISM: bubbles - Galaxies: ISM - X-rays: ISM}

\section{Introduction}
The term ``bubble'' in science is not well defined. A convenient
operational definition could be that of a closed two-dimensional
surface in three-space, separating two media of different (physical)
properties, with lower density inside than outside. Consequently
there exists a vast range of topics in the literature from electron
bubbles in superfluid helium to bubbles in avalanches. In
astrophysics, interstellar bubbles are an already classical subject
in ISM research, and seem to have experienced a renaissance each
time a new process for significant energy injection has been found.
After the discovery of the Stroemgren (1939) sphere, generated by
stellar Lyman continuum photons, it became clear that the rise in
temperature in the H{\sc ii} region, $T_{II}$, with respect to the
neutral ambient medium, $T_{I}$, would imply a strong pressure
imbalance of the order $\sim T_{II}/T_{I}$, and thus a shock wave
would be driven outwards (Oort 1954). The major effect is the growth
of the H{\sc ii} region in size due to the decrease in electron
density $n_e$ inside and hence in recombination rate ($\propto
n_e^2$), allowing stellar photons to propagate further out. The net
result is a bubble filled with low density ionized gas. Another,
even more powerful energy injection mechanism emerged after the
discovery of P~Cygni profiles in stellar spectra (Morton 1967),
revealing the existence of hypersonic stellar winds smashing into
the ISM at speeds of $\sim 2000$ km/s and a canonical mass loss rate
of $10^{-6} \, \msol/{\rm yr}$, or even higher for Wolf-Rayet stars.
Like in the case of the solar wind this leads to a  two-shock
structure, separated by a contact discontinuity that isolates the
wind from the ISM material, but is in reality unstable to
perturbations and allows mixing and mass loading of the stellar wind
flow. Observationally, these bubbles can be detected in X-rays,
since the temperature behind the inner termination shock rises
according to mass and momentum conservation (energy conservation for
a monatomic gas does not give any new information) to $T \approx
\frac{3}{32} \bar m \frac{V_w^2}{k_B} = 5.4 \times 10^7 \, {\rm K}$,
neglecting ambient thermal pressure with respect to ram pressure.
Under these conditions radiative cooling of the bubble is
negligible, and the bubbles are in their so-called energy driven
phase. Adiabatic cooling due to $p-dV$ work on the surrounding
medium, however, is significant. Sweeping up and compressing ISM gas
slows down the outer shock, which can eventually suffer severe
radiative losses, as the downstream density is considerably higher
than behind the stellar wind termination shock. Moreover the dense
outer shell (or part of it) is photoionized and thus well observable
in the optical. Diffuse soft X-ray emission from stellar wind
bubbles has been observed from NGC~6888 with ASCA SIS (Wrigge et al.
1998) and from S~308 with XMM-{\em Newton} EPIC pn (Chu et al.
2003), revealing temperatures of $1.5 \times 10^6$ K and $8 \times
10^6$ K for NGC~6888, and $1.1 \times 10^6$ K, for S~308. As we
shall see, mass loading must play an important r\^ole for decreasing
the temperature thereby enhancing the cooling and thus the X-ray
emissivity.

Already 70 years ago, Baade \& Zwicky (1934) suggested that stellar
core collapse resulting in a neutron star could release a sufficient
amount of gravitational energy to power a supernova (SN) explosion.
The effect of the kinetic energy of $10^{51}$ erg, which is only
$\sim 1$\% of the released neutrino energy, is dramatic. After a
free expansion phase, in which a shell with mass similar to the
ejecta mass is swept up, the pressure difference between shocked ISM
and ejecta drives a reverse shock into the supernova remnant (SNR),
reheating the ejecta as it propagates inwards. The transfer of
energy to the ISM is now determined by the adiabatic Sedov-Taylor
phase, lasting several ten thousands of years until the cooling time
of the outer shock becomes less than the dynamical time scale, and
the SNR enters the radiative phase before eventually merging with
the ISM. The widespread O{\sc vi} line in the ISM is thought to be
the signature of old SNRs.

Young star clusters, like the four million years old NGC~2244
exciting the Rosette Nebula, can blow holes into the emission
nebulae by the \emph{combined} action of several stellar winds. Since the
discoveries of huge H{\sc i} shells, so-called
supershells, either in the Milky Way (e.g.
Heiles 1979, 1980) or in M~31 (Brinks \& Shane 1984), or by
direct observation of huge X-ray emitting
cavities (e.g.\ Cash et al. 1980) in the Orion, Eridanus or Cygnus
regions, it is understood that O- and B-stars in concert can create
\textit{superbubbles} (SBs). Although stellar winds are the initial
contributors over the first few million years, it is obvious that SN
explosions dwarf their energy input over the SB life time of a few
tens of million years. The SBs range in sizes from a few tens to a few
hundreds of parsecs. The most prominent ones are undoubtedly the
Local Bubble (LB), in which our solar system is immersed, and which
is still not well understood, and the adjacent Loop~I SB (LI); both
will be discussed in some detail in this review.
%
%
Extreme examples of bubbles with respect to their energy input on
galactic scales are those driven by (nuclear) starbursts -
often called superwinds -, which in case of NGC~3079  (Cecil et al.\, 2002)
show clear signatures of an outflowing bubble both with Chandra and
HST. Although driven by star formation processes, but for lack of space,
we will discuss neither planetary nebulae, which exhibit a prominent
white dwarf blown wind bubble, nor pulsar wind bubbles driven by
energetic particles, nor bipolar outflows and jets, which show some
additional features such as Mach disks.
Why is it important to study bubbles? Apart from being interesting
astrophysical objects, they are part of the interstellar matter
cycle, enriching the ISM with metals, and, most importantly, they
are the major energy sources of the ISM, controlling its structure
and evolution, with SNRs and SBs being the major contributors.

In Section~\ref{obs} we present some recent observations of the LB
and the LI SB, taken in X-rays and in H{\sc i}. Then, in
Section~\ref{anal} some analytical work, mainly similarity
solutions, and their limitations are discussed, and  in
Section~\ref{num} numerical high resolution simulations of the LB
and LI are shown, finishing off with our conclusions in
Section~\ref{conc}.

\section{Observation of nearby superbubbles}
\label{obs}
\begin{figure}[htbp]
\psfig{file=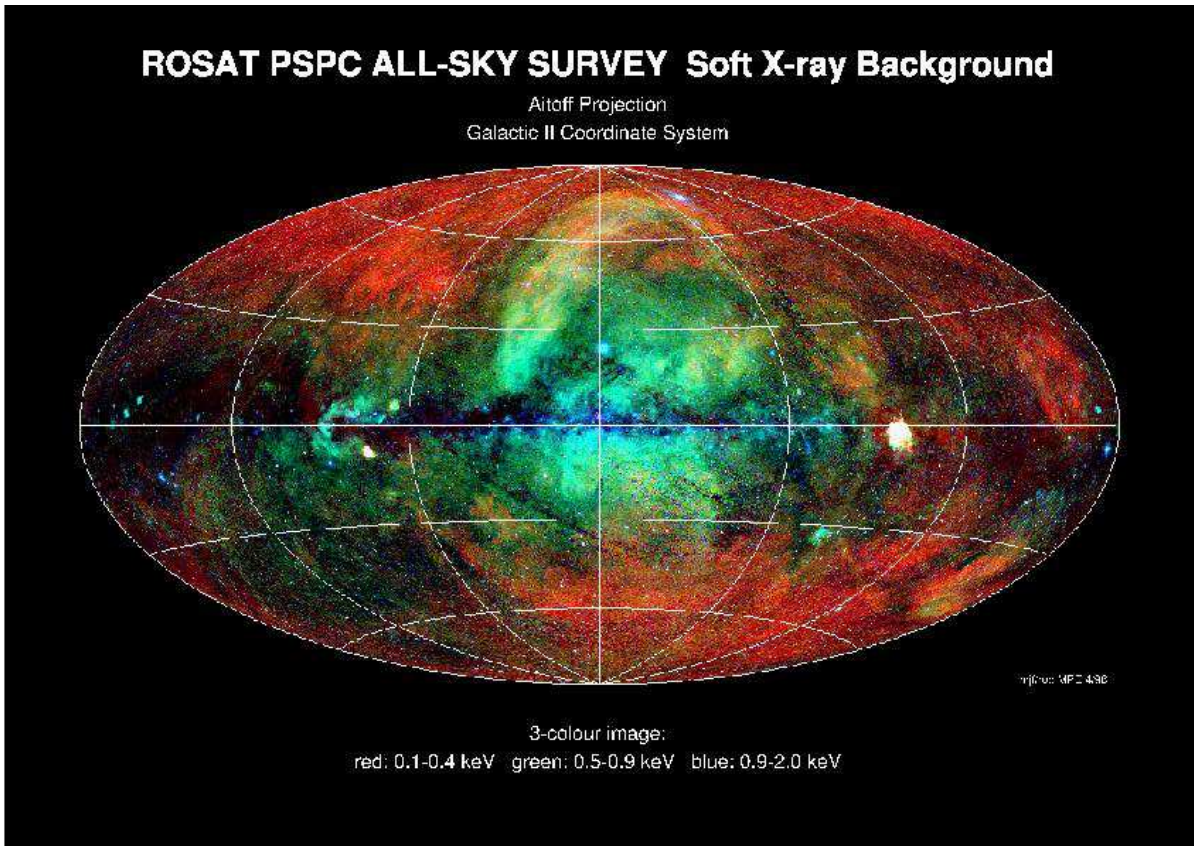,width=0.7\textwidth,angle=270.0,clip=}
\caption{Multispectral X-ray view of the soft X-ray background as
seen by the ROSAT PSPC. The RGB image covers three energy bands
(1/4, 3/4 and 1.5 keV, respectively), and is shown in galactic
coordinates in Aitoff-Hammer projection. The Galactic Centre is at
$l=0^\circ$ with longitudes increasing to the left. The Loop I
bubble lies in the direction of the Galactic Centre and is the
largest coherent X-ray structure in the sky. The North Polar Spur
stretching from $(l,b) = (30^\circ,15^\circ)$ to
$(300^\circ,80^\circ)$ is clearly visible. Since we are located
inside the Local Bubble, we can observe its X-ray emission from all
directions. The picture is taken from Freyberg \& Egger (1999).
 }
\label{rass_rgb}
%
\end{figure}

Since we want to describe bubbles in the \emph{young
local universe}, the closest examples are undoubtedly the LB, in
which our solar system is immersed, and the neighbouring LI SB,
whose outer shell is most likely in contact with the LB shell (cf.\
Egger \& Aschenbach 1995). The centre of the LI bubble is
approximately 250 pc away, and the bubble radius is about 170 pc.
Its proximity is most impressively seen in a ROSAT All Sky Survey
(RASS) multispectral image (Freyberg \& Egger 1999), which shows it
as the largest coherent X-ray structure in the sky (see
Fig.~\ref{rass_rgb}), centred roughly on the Galactic Centre
direction and covering a solid angle of $7/6 \, \pi$ (Breitschwerdt
et al. 1996).
Although the existence of local X-ray emission was realized already
soon after the observation of the diffuse soft X-ray background (SXRB)
(Bowyer et al. 1968), the idea received considerable support by
H{\sc i} observations, revealing a local \textit{cavity} (e.g.\ Frisch \& York,
1983).
\begin{figure}[htbp]
\psfig{file=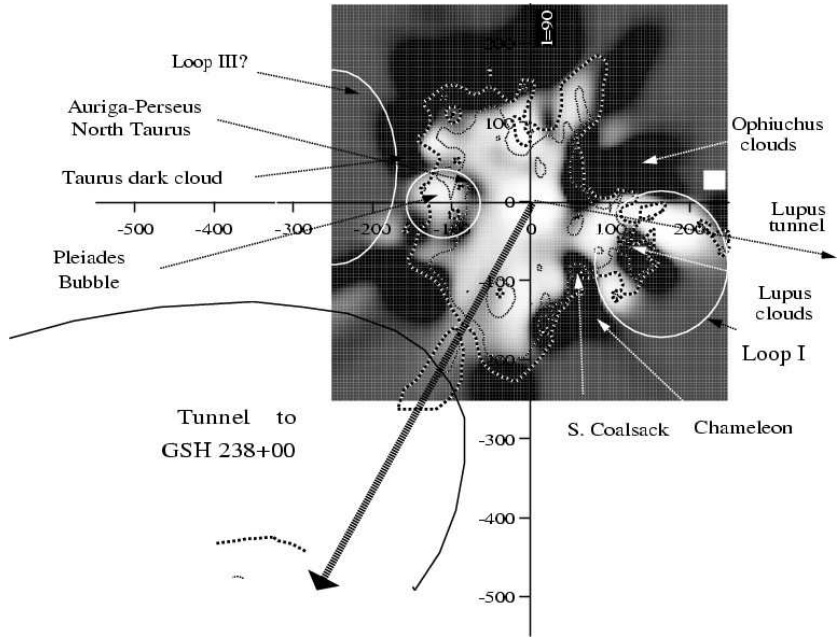,width=0.75
\hsize,bbllx=13pt,bblly=13pt,bburx=215pt,bbury=165pt,clip=}
\caption{Projection of the Local Cavity onto the Galactic plane
inferred from Na{\sc i} absorption line studies by Lallement et al.\
(2003), using sightlines to 1005 stars with known Hipparcos
parallaxes. The grey scale represents the density contrast with
white being low and dark high densities. The dashed lines give the
contours of Na{\sc i} equivalent widths of 20 and $50 \, {\rm
m\AA}$, respectively, with $20$ m{\rm \AA} \, corresponding to
$\log$(Na{\sc i}) = $11.0$ or $N_{\rm H} \approx 2 \times 10^{19} \,
{\rm cm}^{-2}$. The existence of a shell in most directions is
clearly seen, since the small difference between the contours
indicates a steep gradient.}

\label{lb-na1}
\end{figure}
%
The simplest explanation is given by the so-called ``displacement''
or Local Hot Bubble model (Sanders et al., 1977; Tanaka \& Bleeker,
1977), in which it is assumed that the solar system is immersed in a
bubble of diffuse hot plasma in collisional ionization equilibrium,
that displaces H{\sc i}, and has an average radius of $100$ pc. More
recent observations of the Local Cavity (see Fig.~\ref{lb-na1}),
using Na{\sc i} as a sensitive tracer of H{\sc i} (Lallement et al.
2003), show a more complex 3D structure of the hydrogen deficient
hole in the Galactic disk, as well as a clear indication of opening
up into the halo like a chimney. As a natural result, also the X-ray
brightness of the bubble will vary with direction, especially if
entrained clouds are shocked and evaporated. Indeed, observations of
the SXRB exhibit a distinct patchiness in emission, as has been
reported from ROSAT PSPC observations (Snowden et al.\ 2000), and
more recently by a mosaic of observations of the Ophiuchus cloud
(Mendes et al. 2005), which was used to shadow the SXRB and thus
allow to disentangle cloud fore- and background emission.
\begin{figure}[htb]
\centerline{\includegraphics[angle=-90,width=\textwidth,bb=29 28 152 215,clip]{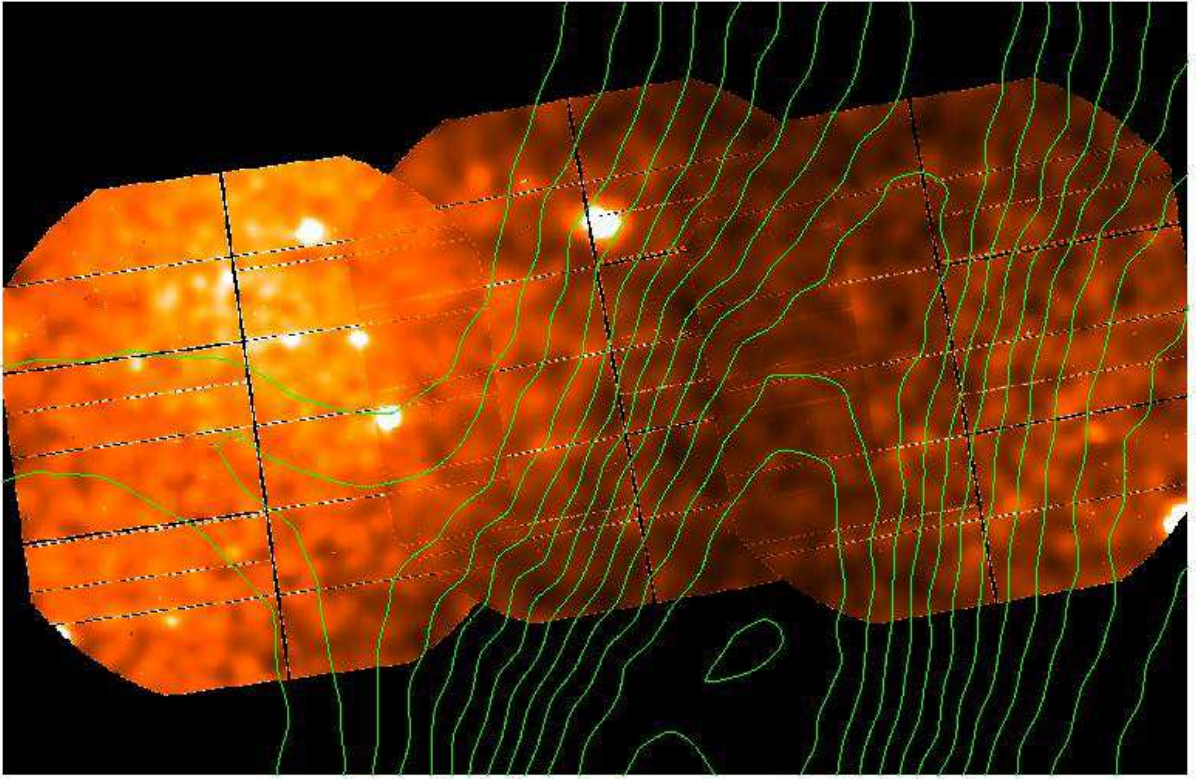}} \caption{Three
individual pointings of the Ophiuchus molecular cloud merged into
one EPIC-pn image, showing the first X-ray shadow detected with
XMM-{\em Newton}, in the energy range $0.5 - 0.9$ keV. There is an
excellent anticorrelation between soft X-ray emission and the
overlaid IRAS 100$\,\mu$ contours (green). The color coding
represents the X-ray intensity with white being the maximum.
} \label{oph-im}
\end{figure}
The absorbing column density in direction to the Ophiuchus cloud,
which is a well-known star forming region at a distance of 150 pc in
direction of the Galactic Centre, contains up to $10^{22} {\rm
cm}^{-2}$ H atoms, efficiently blocking out background radiation up
to 1 keV. This is demonstrated nicely by a deep shadow in diffuse
X-rays, which correlates very well with the IRAS 100 $\mu$ contours
(s.~Fig.~\ref{oph-im}). Our analysis of the spectral composition of
the fore- and background radiation (s.~Fig.~\ref{oph-sp}) shows
convincingly that the major fraction of the emission below 0.3 keV
(most likely unresolved carbon lines) is generated in the
foreground. In addition we observe a significant local fraction of
oxygen lines (O{\sc vii} and O{\sc viii}) between 0.5 and 0.7 keV,
as well as lines between 0.7-0.9 keV (probably iron). These results
give strong observational evidence that the gas inside the LB
is \emph{not} in collisional ionization equilibrium, in
disagreement with the classical Local Hot Bubble model. A small
fraction of the foreground emission stems also from LI.
However, Fig.~\ref{oph-sp} shows that the off-cloud
spectrum contains iron lines, which are absent in the on-cloud
spectrum. Therefore the excitation temperature in the LI
SB must be significantly higher than in the LB,
allowing to disentangle spectrally the respective contributions.
This is not surprising as LI is still an active SB in contrast to
the LB, as we shall see in section~\ref{num}.
\begin{figure}[htb]
\centerline{\includegraphics[angle=-90,width=0.8\hsize,clip]{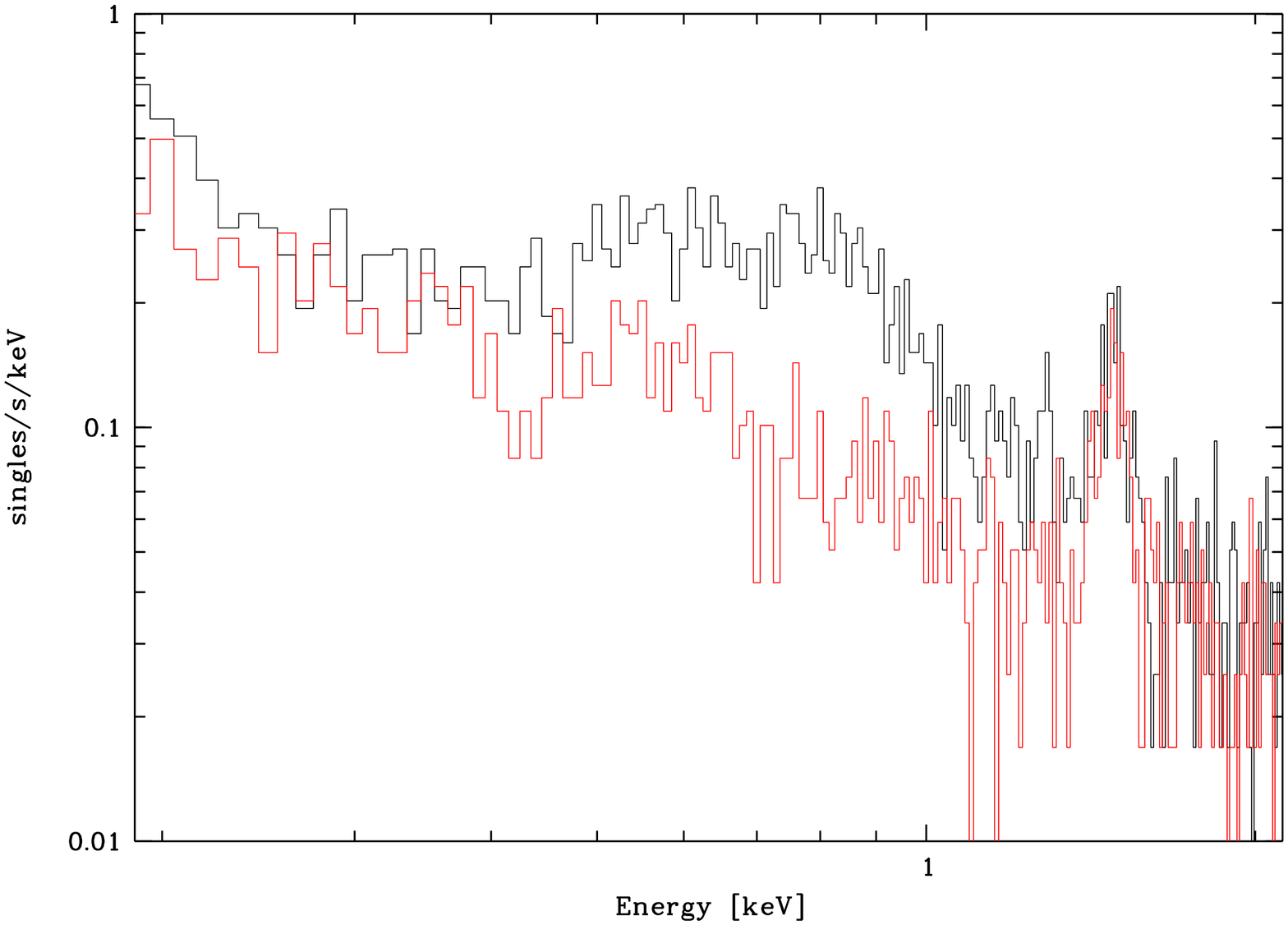}}
\caption{Spectrum of the soft X-ray background towards the Ophiuchus
cloud. We have analyzed XMM-{\em Newton} EPIC pn data from two pointings
of 20 ksec exposure each. Emission line complexes are clearly
distinguishable at $0.5 - 0.7$, $\sim 0.9$ keV, and to a minor
extent at $\sim 0.3$ keV. The on-cloud pointing (red) contains
mainly contributions from the Local Bubble, while the off-cloud
observation has significant contributions from the ambient Loop I
superbubble, as can be seen from the substantial amount of emission
at $0.8 - 0.9$ keV. This can be attributed to iron line complexes,
indicating a higher plasma temperature in Loop I than in the Local
Bubble.
 }
\label{oph-sp}
%
\end{figure}
%
%

The spectral interpretation of the SXRB is far from trivial. First
of all it is at present unclear to what extent different sources
contribute. These are: (i) diffuse local emission from the LB (and
possibly the LI SB, although its major component has a distinctly
higher temperature), (ii) diffuse Galactic emission from the hot ISM
(and other SBs) and unresolved point sources (e.g.\ X-ray binaries),
(iii) a diffuse Galactic halo component (presumably from the
Galactic fountain/wind), (iv) a diffuse extragalactic component
(thought to consist of the WHIM = Warm Hot Intergalactic Medium and
unresolved point sources). Shadowing the darkest regions of the
Milky Way, i.e.\ nearby Bok globules with extinctions of $A_V \sim
30 - 50$ mag, give unmistakably in case of Barnard~68 \emph{two
temperature components} of the local emission (and thus
\textit{inconsistent} with the standard Local Hot Bubble model!): $k
T_1 \approx 0.14\pm 0.04$ keV, and $T_2 \approx 0.20\pm 0.06$ keV
(Freyberg et al. 2004). How is this possible? Several (not mutually
exclusive) explanations have to be further investigated: (i) the LB
is an old SB emitting X-rays from a gas not in ionization
equilibrium (cf.\ Breitschwerdt \& Schmutzler 1994) thus mimicking a
multi-temperature plasma, (ii) there is a significant contribution
from heliospheric plasma which undergoes charge exchange reactions
with highly ionized solar wind atoms (Lallement 2004). At present it
is unclear what the quantitative contribution of the latter process
is (values $\leq 75$\% in the disk and $\leq 50$\% in the halo have
been advocated). Such a very local emission should in principle
exhibit seasonal variations. We have obtained two exposures of the
Ophiuchus cloud, which partially overlap and are 6 months apart. The
differences in emission measure and the spectrum are within the
noise level. Although this is no counterargument it does not support
the hypothesis of a substantial time-dependent variation of the
heliospheric contribution. Further studies are needed to pin down
this crucial component.

\section{Analytical treatment of superbubble (SB) evolution}
\label{anal} The dynamics of SBs has been worked out analytically by
McCray \& Kafatos (1987), based on earlier work by Pikel'ner (1968),
Dyson \& deVries (1972), Weaver et al. (1977) on stellar winds.

A basic principle, which is used in aerodynamics for constructing
models in the wind channel, is the scaling of hydrodynamic flows if
there are no specific length or time scales entering the problem.
Strictly speaking this is never fulfilled, because there are always
boundary layers or time-dependent changes in the flow, but for
studying the large-scale asymptotic behaviour of the flow this
ansatz works remarkably well.  If we e.g.\ neglect the
initial switch-on phase of  a SB, if we assume that the stellar
source region is much smaller than the bubble, and if the
discontinuous energy supply during the SN explosion phase can be
approximated by a continuous injection of mass, momentum and energy,
then \textit{similarity solutions} are reasonably well applicable.

Mathematically speaking, the transformation to a similarity variable
$\xi = (r/A) t^{-\alpha}$, projects the family of solutions of a PDE
system to a one-dimensional family, with all hydrodynamic variables
depending only on the dimensionless similarity variable $\xi$. A
flow is said to be self-similar if its properties at any point $x_1$
and instance of time $t_1$ can be recovered by a similarity
transformation at some other point in spacetime $(x_0, t_0)$. The
exponent $\alpha$ can already be derived from dimensional analysis.
The physical quantities determining the SB dynamics are the energy
injection rate, $L_{\rm SB}$, (with mass and momentum injection
being negligible with respect to the shell mass and momentum during
the energy driven phase) and the ambient density, $\rho_0$. Note
that it is implicitly assumed that the pressure of the ambient
medium can be neglected with respect to the interior pressure of the
bubble. This is certainly valid until the shock becomes weak, in
which case counterpressure has to be included. Then, $\xi =
\left(L_{\rm SB}/\rho_0\right)^{-1/5} r t^{-3/5}$, is the only
possibility to form a dimensionless quantity. Using this similarity
variable, it is now possible to construct the complete flow
solutions in terms of variables $u^\prime(\xi), \rho^\prime(\xi)$
and $P^\prime(\xi)$, obeying matching conditions for boundaries in
the flow at which these variables change discontinuously, like at
the termination shock (where the ``wind'' ejecta are decelerated),
the contact discontinuity (separating the wind from the ISM flow),
and the outer shock (propagating into the ISM).  The integration of
the resulting ODE system is a straightforward but tedious exercise
that can be carried out with the help of an integral, representing
the conservation of the total energy of the system. We can simplify
the procedure considerably by making a few additional, but well justified, assumptions
about the flow in the different regions. Firstly, the ejecta gas,
having a high kinetic energy, is compressed and heated by the strong
termination shock, converting 3/4 of its initial bulk motion into
heat. Therefore the temperature and the speed of sound in this
bubble region are so high, that radiative cooling can be neglected
and the pressure remains uniform for long time. On the other hand
the pressure in the swept-up shell is also uniform due to its
thinness, or in more physical terms, because the sound crossing time
is much less than the dynamical time scale. This is because the
outer shock can cool efficiently, as the ISM density is orders of
magnitude higher than the ejecta gas density, if the latter one is
assumed to be smoothly distributed. In essence, we are allowed to
assume spatially constant density and pressure in the wind bubble
and the shell, respectively.

As it turns out, the assumption of constant energy injection rate
$\lsb$ can be relaxed without violating the similarity argument. In
reality we are dealing with an OB association, in which the stars
are distributed according to some initial mass function (IMF) given
by $\Gamma = {d\log\zeta(\log m) \over d\log m}$; $\zeta$ denotes
the number of stars per unit logarithmic mass interval per unit area
with $\Gamma = -1.1 \pm 0.1$ for stars in Galactic OB associations
with masses in excess of 7 $\msol$ (Massey et al. 1995).
This translates into a number $N(m) \, dm$ of stars in the mass interval
$(m, m+dm)$ (calibrated for some mass interval $N_0 = N(m_0)$), i.e.\
$
N(m) dm = N_0 \left(m\over m_0\right)^{\Gamma-1} dm \,.
$
It can be transformed into a time sequence, if we express the
stellar mass by its main sequence life time, $\tau_{\rm ms}$. For
stars within the mass range $7\msol \leq m \leq 30 \, \msol$ this
can be empirically approximated by $\tau_{\rm ms} = 3\times 10^7 \,
(m/[10 \msol])^{-\eta}$ yr (Stothers 1972), with $\eta = 1.6$. Since
this defines $m$ as a function of time $\tau$, implicitly assuming
that the energy input can be described as a continuous process, we
obtain ${m(\tau)} ={\rm M}_\odot \, \left(\tau \over C\right)^{-1/\eta}
$\,,
with $C = 3.762 \times 10^{16}$ s.

Let then $L_{\rm SB}(t)$ be the energy input per unit time due to a
number of successive SN explosions with a constant energy input of
$E_{\rm SN} = 10^{51}$ erg each, so that the cumulative number of
SNe between stellar masses $m$ and $m_{\rm max}$ reads
\begin{eqnarray}
\tilde N_{\rm SN}(m) = \int_m^{\rm mmax} N(m^\prime ) dm^\prime  =
\frac{N_0 \, m_0}{\Gamma}\left[\left(m^\prime \over
m_0\right)^\Gamma\right]_{m}^{\rm mmax} \,.
\end{eqnarray}
Then we have
\begin{eqnarray}
L_{\rm SB} &=& {d \over dt} (\tilde N_{\rm SN} E_{\rm SN})= E_{\rm SN}
{d \tilde N_{\rm SN} \over dt}
= E_{\rm SN} {d \tilde N_{\rm SN} \over dm} {dm \over d\tau} {d\tau \over dt}\\
&=&\frac{N_0 E_{\rm SN} \msol  \, K^{1-\Gamma}}{\eta C} \,
\left(\frac{\tau_0+t}{C}\right)^{-(\Gamma/\eta + 1)} \,,
\label{eninp1}
\end{eqnarray}
using the previous equations, and putting $m_0 = K \, \msol$.
Since $\tau = t + \tau_0$, where $t$ is the time elapsed since the
first explosion, i.e.\ $\tau_0 = \tau_{\rm MS}(m_{\rm max})$, we
have $d\tau/dt=1$. With the above values for $\Gamma$ and $\eta$, we
obtain the useful formula $L_{\rm SB} = L_0 \, t_7^{\delta}$, where
$\delta = -(\Gamma/\eta + 1) = -0.3125$ and $t_7 = t/10^7$ yr. $L_0$
depends on the richness of the stellar cluster. Thus we see that,
depending on the stellar IMF, the energy input rate by SN explosions
is a mildly decreasing function of time. Although the number of core
collapse SNe increases as the higher masses of the cluster become
depopulated, the increasing time interval between explosions more
than compensates this effect.

If the ambient medium is further assumed to have a constant ambient
density, or one which varies with distance like $\rho \propto
r^{-\beta}$, in which case the similarity variable has to be
transformed to $\alpha = 3/(5-\beta)$, the system can be cast into
the following form:
\begin{equation}
M_{\rm sh}(r) = \int_0^r \rho(r^\prime) d^3 r^\prime \,, \quad
E_{\rm th}(r) = 1/(\gamma -1) \int_0^r p(r^\prime)  d^3 r^\prime \,.
\label{mascon1}
\end{equation}
and the energy input is
shared between kinetic and thermal energy.
Using $\gamma = 5/3$ for the ratio of specific heats, observing that
the bubble pressure $P_b$ remains uniform, and applying spherical
symmetry, conservation of momentum and energy
\begin{equation}
{d\over dt} (M_{\rm sh} \dot R_b) = 4 \pi R_b^2 P_b \,, \quad
{d E_{\rm th} \over dt} = L_{\rm SB}(t) - 4 \pi R_b^2 \dot R_b P_b \,,
\label{mom}
\end{equation}
yields the solution
\begin{eqnarray}
R_b &=& A t^\alpha \,; \quad \alpha = {\delta + 3 \over 5 - \beta} \,,\\
A &=& \left\{{(5-\beta)^3 (3-\beta) \over (7 \delta - \beta - \delta
\beta + 11) (4 \delta + 7 - \delta \beta - 2
\beta)}\right\}^{1/(5-\beta)} \times \left\{{L_0 \over 2\pi (\delta
+ 3) \rho_0} \right\}^{1/(5-\beta)} \,.
\label{simsol1}
\end{eqnarray}
Since the swept-up shell is usually thin, the bubble and shell radius
can be treated as equal during the energy driven phase and are denoted by
$R_b$.
The similarity variable in the case considered here is given by
$\alpha = (2-\Gamma/\eta)/(5-\beta)$ .
For simplicity, the ambient density is assumed to be constant
($\beta=0$), although, as we shall see in our numerical simulations,
this assumption becomes increasingly worse with time. On scales of
ten parsec, the ISM cannot be assumed to be homogeneous any more. As
the cold and warm neutral media are observed to be rather
filamentary in structure, high pressure flows will be channelled
through regions of low density and pressure. It should be mentioned
here, that it is not only the pressure difference between the bubble
and the ambient medium that determines the expansion, as it is
sometimes argued, but also the \textit{inertia} of the shell is a
crucial factor (see~Eq.~\ref{mom}). Therefore mass loading of the
flow is an important factor. Unfortunately, some convenient
assumptions, like e.g.\ the bubble behaves isobaric, do not hold any
more. Pittard et al. (2001a,b) have shown that similarity flow can
be maintained provided the mass loading rate scales as $\dot \rho
\propto r^{(5-7\beta)/3}$ in case of conductive evaporation or $\dot
\rho \propto r^{(-2\beta-5)/3}$ for hydrodynamic mixing according to
the Bernoulli effect. It was assumed that in the former case clumps
passed through the outer shock as it expanded into a clumpy medium
and evaporated in the hot bubble, whereas in the latter case the
clumps were thought to be ejected by the central source itself. Here
it is possible for strong mass loading that the wind flow is slowed
down considerably due to mass pick-up, and in the extreme case even
a termination shock transition can be avoided.

Bergh\"ofer \& Breitschwerdt (2002) have studied the evolution of
the LB under the assumption that 20 SNe from the Pleiades moving
subgroup B1 exploded according to their main sequence life times
with masses between 20 and 10 ${\rm M}_{\odot}$. Using the above
similarity solutions, the radius and the expansion velocity of the bubble evolve as
\begin{equation}
R_b = 251 \left(2 \times 10^{-24} {\rm g}/{\rm cm}^3 \over
\rho_0\right)^{1/5}
      t_7^{0.5375} \, {\rm pc} \,,
\dot R_b = 13.22 \left(2 \times 10^{-24} {\rm g}/{\rm cm}^3
           \over \rho_0\right)^{1/5} t_7^{-0.4625}\, {\rm km/s} \,.
\label{bubrad1}
\end{equation}
As a result of a decreasing energy input rate the exponent in the
expansion law of the radius, $\alpha= 43/80=0.5375$, in
Eq.~(\ref{bubrad1}) is between a Sedov ($\mu=0.4$) and a stellar
wind ($\mu=0.6$) type solution. Thus the present radius of the LB
will be 289 pc and 158 pc and its velocity is 11.7 km/s and 6.4
km/s, if the ambient density is $\rho_0 = 2 \times 10^{-24} \, {\rm
g}/{\rm cm}^3$ and $\rho_0 = 4 \times 10^{-23} \, {\rm g}/{\rm
cm}^3$, respectively (for details see Bergh\"ofer \& Breitschwerdt
2002). In the latter case the value of the ambient density would
correspond roughly to that of the cold neutral medium.
There are several reasons why we may have overestimated the size of
the LB in the similarity solutions above. Firstly, the mass inside
the bubble is significantly higher than the pure ejecta mass, as can
be inferred from the ROSAT X-ray emission measures; when assuming
bubble parameters of $R_b = 100$ pc and $n_b = 5 \times 10^{-3} \,
{\rm cm}^{-3}$ (e.g.\ Snowden et al. 1990) a mass of at least 600
${\rm M}_\odot$ is derived, and using non-equilibrium ionization
plasma models (Breitschwerdt \& Schmutzler 1994) it is even more
than a factor of five higher.
The contribution of ejecta is only of the order of 100 ${\rm
M}_\odot$, and the bulk of the bubble mass is therefore due to
hydrodynamic mixing of shell material, heat conduction between shell
and bubble and evaporation of entrained clouds; hence the flow must
be mass-loaded. The net effect is to reduce the amount of specific
energy per unit mass, because the material mixed in is essentially
cold, thus increasing the rate of radiative cooling. Secondly, the
stellar association has probably been surrounded by a molecular
cloud with a density in excess of $n_0 = 100 \, {\rm cm}^{-3}$ with
subsequent break-out of the bubble and dispersal of the parent cloud
(Breitschwerdt et al. 1996). Thirdly, the number of SN explosions
could be less; here we have assumed that all 20 SNe have occurred
inside the LB. This need not be the case as the subgroup B1 does not
move through the centre of the LB. ROSAT PSPC observations have
revealed an annular shadow centered toward the direction ($l_{\rm
II} = 335^\circ$, $b_{\rm II} = 0^\circ$), which has been
interpreted as an interaction between the LB and the neighbouring LI
SB (Egger \& Aschenbach 1995). The trajectory of the cluster B1 may
have partly crossed the LI region. Alternatively and more likely,
part of the thermal energy might have been liberated into the
Galactic halo, since there is some evidence that the LB is open
toward the North Galactic Pole (see Lallement et al. 2003). It
should also be mentioned that due to small number statistics the
true number of SNe can vary by a factor of 2. Finally, although
there is no stringent evidence, it would be very unusual, if the LB
would not be bounded by a magnetic field, whose tension and pressure
forces would decrease the size of the LB.

Given these uncertainties and the fact that the simple analytic
model discussed above can only be considered as an upper limit, the
direct comparison with observations is not convincing. The bubble
radius and shell velocity are rather insensitive to the energy input
rate and the ambient density (due to the power of $1/5$) and
therefore not well constrained, but depend more sensitively on the
expansion time scale. Thus we can only assert with some confidence
that the age of the LB should be between $1 - 2 \times 10^7$ yr.

The most serious drawback of analytical solutions in general and of
similarity solutions in particular, is the assumption of homogeneity
of the ambient medium on scales exceeding about 10 pc. To see this,
consider the area coverage of the disk by hot gas, which is
$\xi_{\rm SN} \sim \nu_0/2 \, \tau_{\rm SN}\, (R_{\rm SN}/R_{\rm
gal})^2 \approx 0.67$ due to SNe, and $\xi_{\rm SB} \sim \nu_0/2 \,
\tau_{\rm SB}\, (R_{\rm SB}/R_{\rm gal})^2 \approx 0.9$ due to SBs,
respectively, assuming that half of the explosions go off randomly,
and half in a clustered fashion within a star forming disk of 10 kpc
radius and a disk SN rate of $\nu_0 = 2$ per century. Here we used
the final SN radius according to McKee \& Ostriker (1977), being
$R_{\rm SN}\approx 55$ pc after $\tau_{\rm SN} \approx 2.2 \times
10^6$ yr and the SB radius from the paper of McCray \& Kafatos
(1987) of $R_{\rm SB} \approx 212$ pc after $\tau_{\rm SB} \approx
10^7$ yr, for a typical cluster with 50 OB stars. The overturning
rate of a typical patch of ISM will then roughly be between $3.4
\times 10^6$ yr and $1.1 \times 10^7$ yr for SNe and SBs,
respectively. Since we did not take into account overlapping of
remnants and bubbles these values are lower limits. As will be shown
in the next section, a roughly constant star formation rate and
hence SN rate for a Galactic initial mass function (IMF) will lead
to an ISM background medium that is highly irregular in density and
temperature (and even pressure variations within an order of
magnitude are observed) and it bears a high level of turbulence.

\section{Numerical simulations of the Local Bubble (LB) and
Loop~I (LI) evolution}
\label{num}
We have performed high resolution 3D simulations of the Galactic
disk and halo (Avillez \& Breitschwerdt 2004, 2005; see also this volume)
on a grid of $1\, {\rm kpc} \times 1 \, {\rm kpc}$ in the plane and $z=\pm 10 \,
{\rm kpc}$ perpendicular to it. Using AMR technique, we obtained
resolution of scales down to 1.25 pc for MHD, and 0.625 pc for pure
hydrodynamical (HD) runs.
These simulations, which revealed many new
features of the ISM, e.g.\ low volume filling factor of hot gas in
the disk, establishment of the fountain flow even in the presence of
a disk parallel magnetic field, more than half of the mass in
classical thermally unstable regions, serve as a \emph{realistic
background medium} for the expansion of the LB and the LI SB.
We took data cubes of HD runs and picked up a site
%
\begin{figure}[htbp]
\centering
\includegraphics[width=0.8\hsize,bb=20 18 106 106,angle=0,clip]{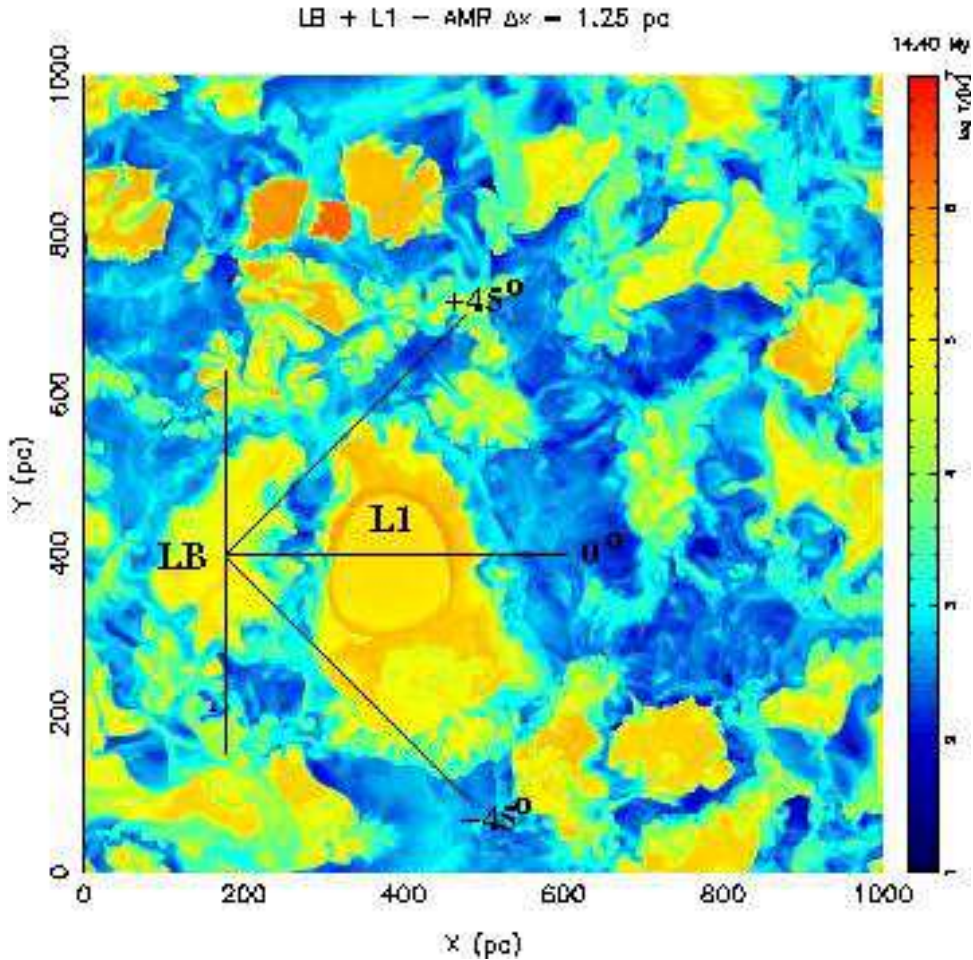}
\caption{Temperature map (cut through Galactic plane)
    of a 3D Local Bubble simulation, 14.4 Myr after the first
    explosion; LB is centered at (175, 400) pc and Loop I
    at (375, 400) pc.}

\label{fig_temp}
\end{figure}
with enough mass to form the 81 stars, with masses, $M_*$, between 7
and 31 ${\rm M}_{\odot}$, that represent the Sco-Cen cluster inside
the LI SB; 39 massive stars with $14 \leq M_* \leq 31 \, {\rm
M}_{\odot}$ have already exploded, generating the LI cavity. At
present the Sco-Cen cluster (arbitrarily located at $(375,400)$ pc
has 42 stars to explode within the next 13 Myrs). We followed the
trajectory of the moving subgroup B1 of Pleiades (see~ Bergh\"ofer
\& Breitschwerdt 2002), whose SNe in the LB went off along a path
crossing the solar neighbourhood.
%
As a result, we observe that the locally enhanced SN rates produce
coherent LB and LI structures (due to ongoing star formation) within
a highly disturbed background medium (see Fig.~\ref{fig_temp}). The
successive explosions heat and pressurize the LB, which at first
looks smooth, but develops internal temperature and density
structure at later stages. After 14 Myr the 20 SNe that occurred
inside the LB fill a volume roughly corresponding to the present day
size (see Fig.~\ref{fig_temp}, bubbles are labelled by LB and L1).
The cavity is still bounded by an outer shell, which exhibits holes
due to Rayleigh-Taylor instabilities, as has been predicted
analytically by Breitschwerdt et al. (2000), and it will start to
fragment in $\sim 3$ Myr from now.
%
\begin{figure}[htbp]
\centering
\includegraphics[width=0.45\hsize,angle=0]{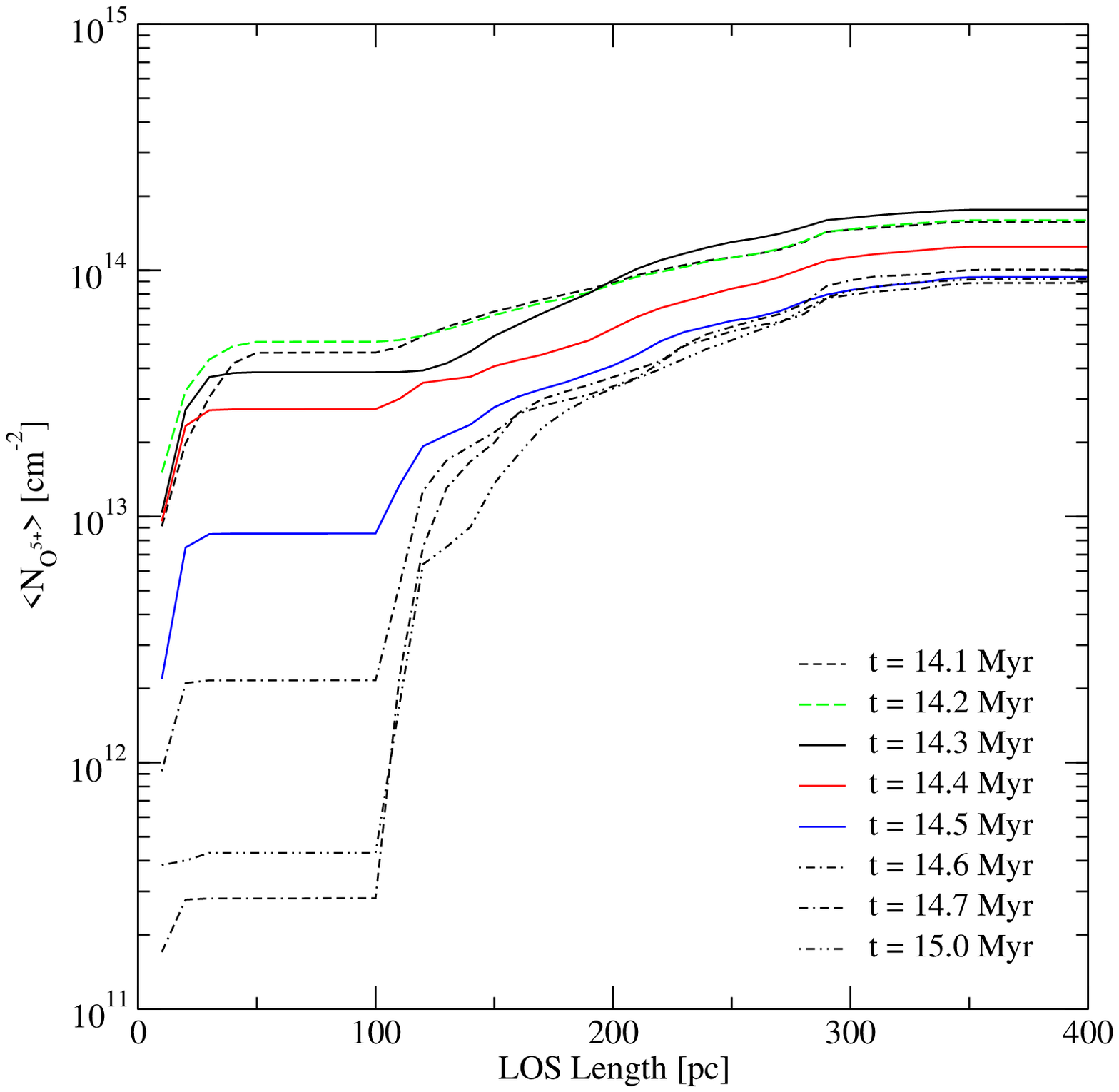}
\includegraphics[width=0.45\hsize,angle=0]{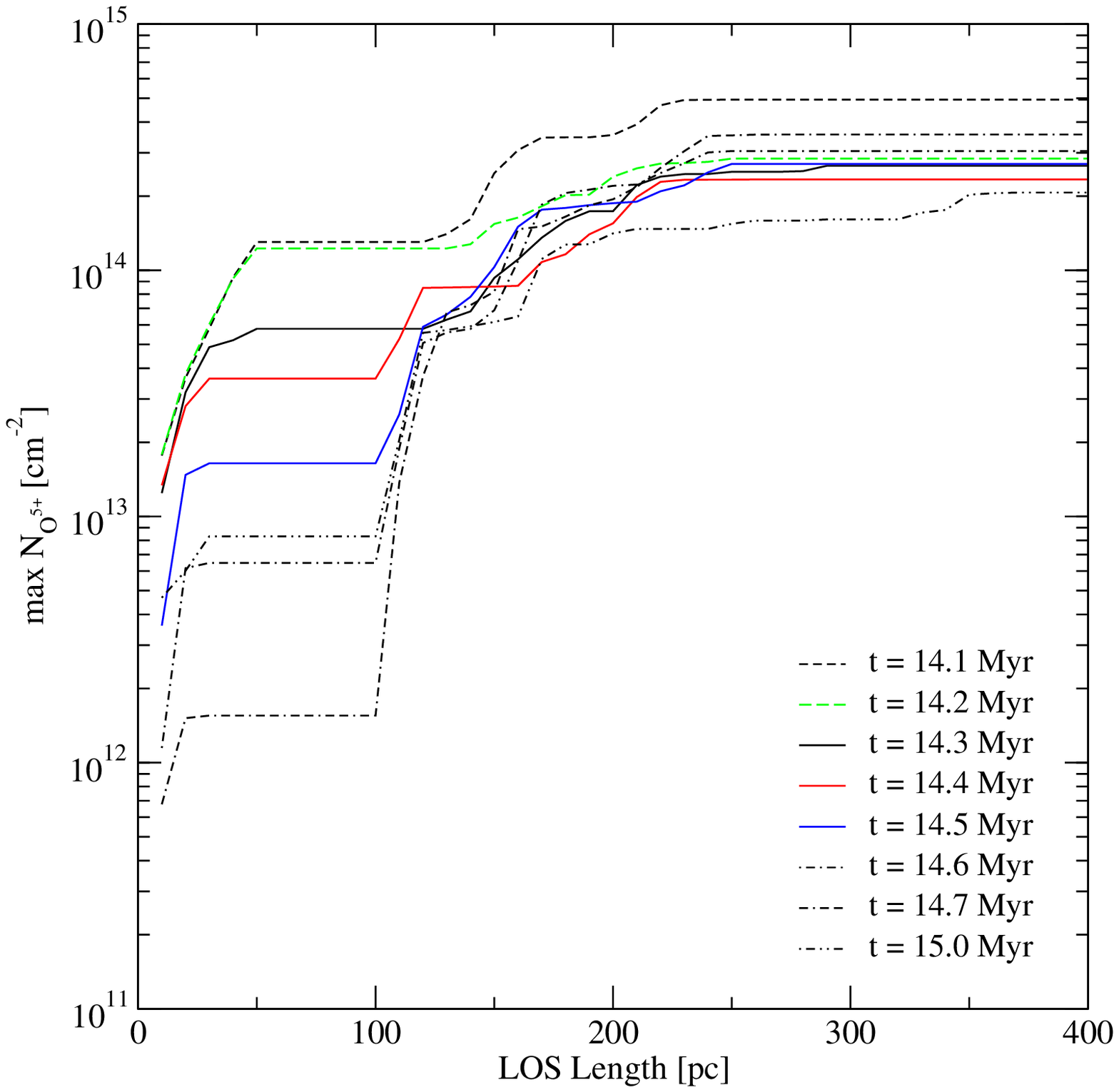}\\
\caption{ O{\sc vi} column density averaged over
  angles (left panel) indicated in Fig.~\ref{fig_temp} and maximum
  column density (right panel) as a function of LOS path length at
  $14.1 \leq t\leq 15$ Myr of Local and Loop I bubbles evolution.}
\label{fig3_ovi}
\end{figure}
It has been argued that a crucial test of any LB model is the column
density of the interstellar ion O{\sc vi} (Cox 2004), whose
discovery back in the 70's led to the establishment of the hot
intercloud medium. So far all models have failed to reproduce the
fairly low O{\sc vi}-value, most recently measured with FUSE
(Oegerle et al. 2004), to be $N_{\rm OVI} \simeq 7\times 10^{12}$
cm$^{-2}$. To compare this with our simulations we have calculated
the average and maximum column densities of O{\sc vi}, i.e.,
$\langle \mbox{N(O{\sc vi})} \rangle$ and
$\mbox{N}_{\mbox{max}}\mbox{(O{\sc vi})} $ along 91 lines of sight
(LOS) extending from the Sun and crossing LI from an angle of
$-45\deg$ to $+45\deg$ (s.~Fig.~\ref{fig_temp}). Within the LB
(i.e., for a LOS length $l_{LOS}\leq 100$ pc) $\langle \mbox{N(O{\sc
vi})} \rangle$ and $\mbox{N}_{\mbox{max}}\mbox{(O{\sc vi})} $
decrease steeply from $5\times10^{13}$ to $3\times 10^{11}$
cm$^{-2}$ and from $1.2\times 10^{14}$ to $1.5\times 10^{12}$
cm$^{-2}$, respectively, for $14.1 \leq t\leq 15$ Myr
(Fig.~\ref{fig3_ovi}), because no further SN explosions occur and
recombination is taking place. For LOS sampling gas from outside the
LB (i.e., $l_{LOS}>100$ pc) $\langle \mbox{N(O{\sc vi})} \rangle >
6\times 10^{12}$ and $\mbox{N}_{\mbox{max}}\mbox{(O{\sc vi})} >
5\times 10^{13}$ cm$^{-2}$. We have made histograms of column
densities obtained in the 91 LOS for $t=14.5$ and 14.6 Myr, which
show that for $t=14.6$ Myr all the LOS have column densities smaller
than $10^{12.9}$ cm$^{-2}$, while for $t=14.5$ Myr 67\% of the lines
have column densities smaller than $10^{13}$ cm$^{-2}$ and in
particular 49\% of the lines have $\mbox{N(O{\sc vi})}\leq 7.9\times
10^{12}$ cm$^{-2}$. Noting that in the present model at 14.5 Myr the
O{\sc vi} column densities are smaller than $1.7\times 10^{13}$
cm$^{-2}$ and $\langle \mbox{N(O{\sc vi})} \rangle = 8.5\times
10^{12}$ cm$^{-2}$ (see the respective lines in both panels of
Fig.~\ref{fig3_ovi}), we are thus able to reproduce the
measured $\langle \mbox{N(O{\sc vi})} \rangle$ values, provided that
the age of the LB is $\sim 14.7^{+0.5}_{-0.2}$ Myrs.

\section{Conclusions}
\label{conc}
Galactic and extragalactic interstellar bubbles are still an active
area of research. Despite the widespread belief that H{\sc ii}
regions, SNRs, stellar wind bubbles and superbubbles are fully
understood in theory, it has to be emphasized that \emph{real
bubbles}, observed in the Galaxy, such as the Local or LI
superbubbles, or in external galaxies, such as in the LMC, are often
poorly fitted by standard similarity solutions. The reason lies in
the inapplicability of major assumptions, e.g., that the ISM is
homogeneous, and that the bubbles are either in an energy or
momentum conserving phase. High resolution 3D simulations in a
highly structured and turbulent background medium offer a much
better description and include physical processes such as mass
loading and turbulent mixing on a fundamental level. Although this
drains heavily on computer resources, the increased precision of
observations in the near future will warrant such an effort.

\section*{Acknowledgments}
DB would like to thank Thierry Montmerle and Almas Chalabaev for
their excellent organization, for financial help and patience with
the manuscript. It was a great pleasure to be in La Thuile. He also
thanks Verena Baumgartner for proofreading of the text.

\end{document}